\documentclass[aps,twocolumn,superscriptaddress,floatfix,showpacs]{revtex4}

\usepackage{graphicx}
\usepackage{dcolumn}
\usepackage{epsfig}
\usepackage{amssymb}
\usepackage{amsmath}
\usepackage{rotating}
\usepackage{color}
\usepackage{subeqnarray}
\newcommand\T{\rule{0pt}{2.6ex}}
\newcommand\B{\rule[-1.2ex]{0pt}{0pt}}
\usepackage[hyperindex,breaklinks]{hyperref}

\begin{document}

\title{Unexpected spatial distribution of bubble rearrangements in coarsening foams}

\author{David A. Sessoms}\affiliation{University of Fribourg, Dept. of Physics, Chemin du Musée 3, CH-1700 Fribourg,
Switzerland}
\author{Hugo Bissig}\affiliation{University of Fribourg, Dept. of Physics, Chemin du Musée 3, CH-1700 Fribourg,
Switzerland}
\author{Agn\`{e}s Duri}
\author{Luca Cipelletti}\affiliation{LCVN UMR 5587, CNRS and University of Montpellier II, 26 Place E. Bataillon,
F-34095 Montpellier, France}
\author{V\'{e}ronique Trappe} \email[Electronic address: ]{Veronique.Trappe@unifr.ch} \affiliation{University of Fribourg, Dept. of Physics, Chemin du Musée 3, CH-1700 Fribourg,
Switzerland}

\date{\today}

\begin{abstract}

Foams are ideal model systems to study stress-driven dynamics, as stress-imbalances within
the system are continuously generated by the coarsening process, which unlike thermal
fluctuations, can be conveniently quantified by optical means. However, the high turbidity of
foams generally hinders the detailed study of the temporal and spatial distribution of
rearrangement events, such that definite assessments regarding their contribution to the
overall dynamics could not be made so far. In this paper, we use novel light scattering
techniques to measure the frequency and position of events within a large sample volume. As
recently reported (A. S. Gittings and D. J. Durian, \textit{Phys.\ Rev.\ E}, 2008, \textbf{78}, 066313),
we find that the foam dynamics is determined by two distinct processes:
intermittent bubble rearrangements of finite duration and a spatially homogeneous quasicontinuous
process. Our experiments show that the convolution of these two processes
determines the age-dependence of the mean dynamics, such that relations between
intermittent rearrangements and coarsening process can not be established by considering
means. By contrast the use of the recently introduced photon correlation imaging technique (A. Duri, D. A. Sessoms, V. Trappe, and L. Cipelletti, \textit{Phys.\ Rev.\ Lett.}, 2009, \textbf{102}, 085702)
enables us to assess that the event frequency is directly determined by the strain-rate imposed
by the coarsening process. Surprisingly, we also find that, although the distribution of
successive events in time is consistent with a random process, the spatial distribution of
successive events is not random: rearrangements are more likely to occur within a recently
rearranged zone. This implies that a topological rearrangement is likely to lead to an unstable
configuration, such that a small amount of coarsening-induced strain is sufficient to trigger
another event.

\end{abstract}


 \maketitle

\section{Introduction}
Foams consist of gas bubbles that are densely packed in a continuous liquid phase. Because of
the high packing fraction, the bubbles are deformed, such that the system is constantly under
internal stress. Though thermal motion is insufficient to lead to any significant restructuring
of the random bubble configuration, foams reconfigure in time because of coarsening: larger
bubbles growing at the expense of the smaller ones due to the difference in their Laplace
pressures. This process leads to imbalances of the internal stresses, which are released by
intermittent, local rearrangements of bubbles~\cite{Weaire1999, DURIAN1991}. Thus, the coarsening process provides the
system with an internal mechanical means of "thermalization" that eventually allows for a
complete reconfiguration of the foam. Such stress-driven relaxation is the hallmark of a
variety of other soft glassy systems~\cite{Sollich1997}, such as colloidal gels~\cite{Cipelletti2000}, concentrated microgel systems~\cite{Sessoms2009},
and surfactant phases~\cite{Ramos2005}; however, in these systems, the mechanism leading to imbalanced
stresses is not well understood. By contrast, in foams the coarsening process is a quantifiable
measure of the temporal evolution of stress-imbalances, and thus foams are ideal benchmark
systems to study stress driven relaxation processes.

Their dynamics has been previously investigated using diffusing-wave spectroscopy
(DWS)~\cite{DURIAN1991, Cohen-Addad2001, Gopal1995, Hoehler1997, Earnshaw1994, Gopal1999, Gopal1997}, an extension of dynamic light scattering to the multiple scattering limit~\cite{Pine1993, Maret1997}.
However, traditional DWS relies on extensive time and space averaging, thereby precluding
direct characterization of the intermittent dynamics. Indeed, recent speckle visibility
experiments revealed that the dynamics of foams exhibits features that were not considered in
previous modelling of the time-averaged correlation functions~\cite{Gittings2008}. In particular, it was shown
that the dynamic light scattering signal was not only determined by intermittent
rearrangements of the local bubble configuration, but that another process of undefined origin
was contributing as well. Such mix of dynamic processes, however, significantly complicates
the interpretation of time and space averaged dynamic light scattering signals that have been
used so far to characterize foam dynamics. In particular, it obscures the evaluation of the
relation between the intermittent restructuring events and the coarsening process, which is
essential to understand the interplay between stress accumulation, local yielding and structural
relaxation.

In this article, we extend the investigations of foam dynamics to spatially resolved
dynamics, applying the recently introduced photon correlation imaging scheme~\cite{Duri2009} to the
backscattering plane of a coarsening foam. To correlate our findings to previous work, we
simultaneously perform time resolved experiments~\cite{Cipelletti2003, Duri2005}, in the classical transmission far field
geometry of DWS. Our results reveal that the event frequency is directly correlated to the
strain imposed by the coarsening process, which is not evident when considering the time and
spatially averaged signals within the frame of previous models, where the existence of an
additional dynamic process was not considered. Moreover, we find that though the occurrence
of events is random in time it is not in space. Indeed, our results suggest that successive
events preferentially occur at the same location.

\section{Experimental}
Our foam is a commercial Gillette shaving foam that consists of polydisperse bubbles tightly
packed in an aqueous solution of stearic acid and triethanolamine. As demonstrated in a
number of investigations~\cite{Gittings2008, DURIAN1991, Earnshaw1994, Gopal1997, Gopal1995, Hoehler1997, Durian1991PRA, Cohen-Addad2001, Cohen-Addad1998, HohlerEPL1999,Cohen-Addad2004}, this type of foam exhibits dynamical and rheological
properties that are remarkably insensitive to variations in the detailed composition, which
inevitably will vary as a function of the market evolution. This insensitivity denotes Gillette
shaving foams as an ideal benchmark system to study different features of foam dynamics,
yielding results that can be compared to the large number of prior investigations.

\begin{figure}[tbp]
\begin{center}
\scalebox{1.1}{\includegraphics{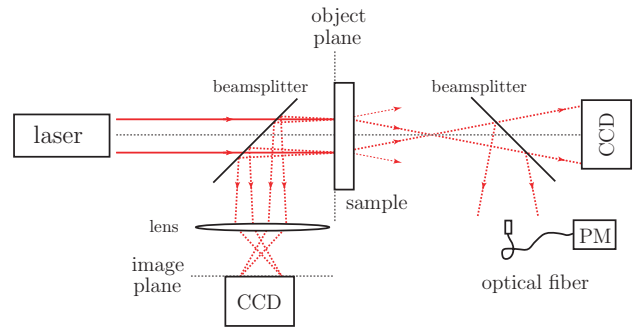}}
\end{center}
\caption{Schematic of light scattering set-up used to record simultaneously
the speckle pattern formed in the far-field transmission geometry and in
the backscattering plane of the sample using CCD-cameras. To assess the
evolution of the bubble size at each run, the transmitted light intensity is
measured continuously using a photo-multiplier (PM) detection scheme.}
\label{fig:schematic}
\end{figure}

To characterize the coarsening process of our system, we perform both optical
microscopy and diffuse light transmittance experiments~\cite{Pine1993}. Immediately after production, the
foam is injected into a rectangular cell of thickness $L = 2$mm or 5mm and kept at a stable
ambient temperature of T = 21$^{\circ}$C. The age of the sample $t_{w}$ is defined as the time elapsed
since sample production. We record the evolution of the bubble size by imaging the foam
surface, where we determine the mean bubble size $\left\langle a_{t_{w}}\right\rangle$ based on the number
of bubbles detected within the area defined by our field of view.  As previously shown,
$\left\langle a_{t_{w}}\right\rangle$ is directly proportional to the transport mean free path $l^{*}$ that we measure in diffuse transmittance experiments~\cite{DURIAN1991,Durian1991PRA}. To perform these experiments we illuminate our sample with a
collimated laser beam with a wavelength of 532~nm \textit{in vacuo} and a diameter of 9~mm. A schematic layout of our experimental set-up is shown in Fig.~\ref{fig:schematic}.
We detect the mean intensity of the transmitted light as a function of $t_{w}$ using a photomultiplier
and determine the age dependent transport mean free path of our sample using the transmitted
intensity of a reference of known $l^{*}$ to calibrate our experiment. These experiments are
performed simultaneously to the diffusing wave spectroscopy experiments outlined below,
which enables us to determine slight variations in the coarsening process in different runs of
the experiment. By mapping the evolution of $l^{*}$ to the evolution of $\left\langle a_{t_{w}}\right\rangle$, we find that for
our foam $l^{*} / \left\langle 2 \cdot a_{t_{w}}\right\rangle = 4.6 \pm 0.3 $, which is somewhat larger than the typical values found in
previous experiments probing the optical properties of Gillette foams~\cite{DURIAN1991,Durian1991PRA}. In agreement with
previous results we find that the bubble size evolution is well described by
$\left\langle a_{t_{w}}\right\rangle^{2} - \left\langle a_{0}\right\rangle^{2} \propto {l^{*}_{t_{w}}}^{2} - {l^{*}_{0}}^{2} \propto t_{w}$, with $\left\langle a_{0}\right\rangle$ and $l^{*}_{0}$ respectively the bubble radius and
the transport mean free path at $t_{w}=0$~\cite{Cohen-Addad2001, Durian1991PRA}.  Our main experiments are performed at two
different ages, $\overline{t_{w}} \cong$ 272 min (age 1) and $\overline{t_{w}} \cong $ 663 min (age 2), where $\overline{t_{w}}$ denotes the age of
the sample at the middle of the experiment. The bubble radii are respectively $\simeq 50 \mathrm{\mu m}$  and
$\simeq 76 \mathrm{\mu m}$ at age 1 and age 2. The duration of our experiments is 800s at age 1 and 1200s at
age 2, which ensures that the bubble size does not change by more than 2\% in the course of
our experiments.

To characterize the dynamics of our foam at age 1 and age 2, we take advantage of the
concepts of diffusing wave spectroscopy. Of particular relevance is here the model that has
been developed to describe a gradual reconfiguration of a foam by intermittent
rearrangements of the local bubble configuration~\cite{DURIAN1991,Durian1991PRA}. In this model, the rearrangements are
presumed to occur randomly in space and time, to be instantaneous and to affect a well
defined localized region of radius $\xi$. Assuming that all scattered light passing a rearranged
region is completely dephased, while photon paths not passing through a rearranged region
remain unchanged, the field autocorrelation $\mathrm{g_{1}}\left(\tau\right)$ can be regarded as a direct measure of the
fraction of paths not rearranged during the lag time $\tau$; for transmission experiments $\mathrm{g_{1}}\left(\tau\right)$ is
then well approximated by
\begin{equation}
\mathrm{g_{1}}\left(\tau\right) \cong \exp \left\{ - R\cdot \tfrac{4}{3}\pi \xi^{3} \cdot \left(L/l^{*}\right)^{2}\cdot \tau\right\}
\label{eq:g1trans}
\end{equation}
with $R$ the rate of intermittent rearrangements per unit volume.  The characteristic decay time
of the field correlation function $\left(R\cdot \tfrac{4}{3}\pi \xi^{3}\right)^{-1}$ thus denotes the mean time between events at a given location. To gain a physical intuition about the impact of a single event on $\mathrm{g_{1}}\left(\tau\right)$ for a
given transmission experiment, we can rewrite equ.~\ref{eq:g1trans} with respect to the actual volume
element observed, the scattering volume $V_{sc}$.
\begin{equation}
\mathrm{g_{1}}\left(\tau\right) \cong \exp \left\{ - \Gamma_{V_{sc}}\cdot \frac{\tfrac{4}{3}\pi \xi^{3}}{V_{sc}} \cdot \left(L/l^{*}\right)^{2}\cdot \tau\right\}
\label{eq:g1transVsc}
\end{equation}
Here, $\Gamma_{V_{sc}}$ is the rate of events occurring within the scattering volume, such that $\Gamma_{V_{sc}} \cdot \tau$
corresponds to the number of events occurring within $V_{sc}$ during $\tau$. The fraction of the paths
randomized per event is $\left( \tfrac{4}{3}\pi \xi^{3} / V_{sc}\right) \cdot \left(L/l^{*}\right)^{2}$; it depends on the fraction of the scattering
volume rearranged per event, $\tfrac{4}{3}\pi \xi^{3}$, and, due to the effective diffusion of the light
through the sample, on $\left(L/l^{*}\right)^{2}$.

Our experiments are based on the classical schemes of dynamic light scattering and on
the use of charge coupled device (CCD) cameras to capture the intensity fluctuations of a
large number of independent speckles simultaneously. We record the evolution of the speckle
pattern in time using two different schemes.

Placing a CCD-detector far from the light-exit-plane of our
sample, as shown in Fig.~\ref{fig:schematic}, we record the speckle pattern in the
far-field transmission geometry and process the images using the
time resolved correlation (TRC) scheme~\cite{Cipelletti2003,Duri2005}. Here, we calculate
a time resolved degree of correlation between two speckle
patterns recorded at time $t_{w}$ and $t_{w}+\tau$, according to
\begin{equation}
c_{I}\left(t_{w},\tau\right)= \frac{1}{\beta}\left( \frac{\left\langle I_{p}\left(t_{w}\right) I_{p}\left(t_{w}+\tau\right)\right\rangle_{p}}{\left\langle I_{p}\left(t_{w}\right) \right\rangle_{p} \left\langle I_{p}\left(t_{w}+\tau\right) \right\rangle_{p}} -1\right)
\label{eq:cI}
\end{equation}
where $I_{p}$ is the intensity recorded at the $p$-th pixel of the CCD
array and $\left\langle \ldots \right\rangle_{p}$ denotes an average over all pixels of the speckle
image; $\beta$ is a prefactor that depends on the speckle-to-pixel size
ratio, chosen so that $\overline{c_{I}\left(t_{w},\tau\right)} = 1$ for $\tau \rightarrow 0$, where $\overline{\cdots}$ denotes
a time average over the duration of the experiments. Note that
the time averaged signal corresponds to the usual intensity
correlation function $\mathrm{g_{2}}\left(\tau\right)-1=\overline{c_{I}\left(t_{w},\tau\right)}$, which relates to the
field correlation function \textit{via} the Siegert relation, $\mathrm{g_{2}}\left(\tau\right)-1 = \mathrm{g_{1}}\left(\tau\right)^{2}$. The TRC-method enables us to access the temporal fluctuations
in the dynamics of our samples with a spatial average
taken over the entire scattering volume. The magnitude of the
fluctuations due to the occurrence of local rearrangement events
will here depend on the effectiveness of a single event to
contribute to a decrease in $c_{I}\left(t_{w},\tau\right)$, which we expect to depend on
the fraction of the paths dephased per event; based on the model
discussed above this corresponds to $\tfrac{4}{3}\pi \xi^{3}/V_{sc}\left(L/l^{*}\right)^{2}$.

\begin{figure}[tbp]
\begin{center}
\scalebox{0.28}{\includegraphics{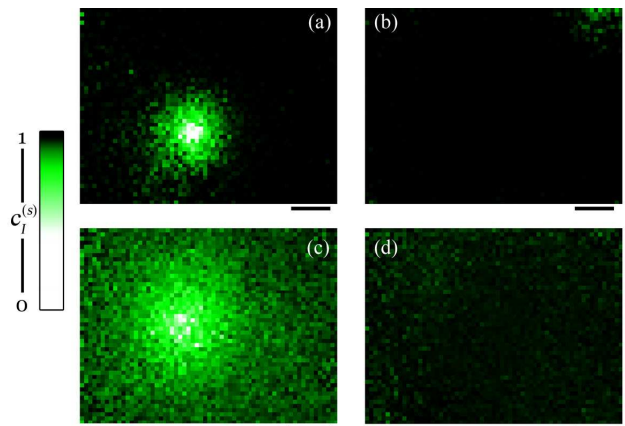}}
\end{center}
\caption{Representative dynamic activity maps depicting respectively
a moment at which an event occurs [(a) age 1 and (c) age 2] and a moment
during the periods between events [(b) age 1 and (d) age 2]. The scale bar
corresponds to 1 mm. Note that the spatial extent of the rearrangement
typically scales with the bubble size.}
\label{fig:dams}
\end{figure}

The second method used is the recently introduced photon
correlation imaging technique~\cite{Duri2009}, which is closely related to other
speckle imaging and near field scattering techniques~\cite{Briers2007,Zakharov2010,Erpelding2008,Cerbino2009}. It
enables us to resolve dynamics with both temporal and spatial
resolution. Instead of recording the speckle pattern in the far
field, where each speckle contains phase contributions of all
scatterers within the sample, we image the scattered intensity
pattern at the entrance plane of the sample onto the camera, as
shown in Fig.~\ref{fig:schematic}, and calculate the instantaneous degree of
correlation with spatial resolution
\begin{equation}
c^{\left(s\right)}_{I}\left(t_{w},\tau\right)= \frac{1}{\beta}\left( \frac{\left\langle I_{p}\left(t_{w}\right) I_{p}\left(t_{w}+\tau\right)\right\rangle_{p\ ROI}}{\left\langle I_{p}\left(t_{w}\right) \right\rangle_{p} \left\langle I_{p}\left(t_{w}+\tau\right) \right\rangle_{p}} -1\right)
\label{eq:cIspace}
\end{equation}
where $\left\langle \ldots \right\rangle_{p\ ROI}$ denotes an average over pixels belonging to
a region of interest (ROI) centred around the position $s$. The
spatially resolved dynamics of the foam at a moment defined by
$t_{w}$ and $t_{w}+\tau$ can then be presented graphically by constructing
dynamical activity maps, where each pixel of the map is colour
coded to represent the local degree of correlation $c^{\left(s\right)}_{I}\left(t_{w},\tau\right)$ characterizing
the fluctuations of diffuse light emanating from
a specific area of the sample. Examples of such dynamical
activity maps are shown in Fig.~\ref{fig:dams}. The occurrence of events is
clearly depicted as a bright patch corresponding to a low $c^{\left(s\right)}_{I}\left(t_{w},\tau\right)$
in the dynamical activity map, as shown for age 1 and age 2 in
Fig.~\ref{fig:dams}(a) and~(c), respectively. 
As our samples are turbid, the
appearance of a spatially resolved event will be limited to some
critical depth within the sample, since the backscattered light is
effectively diffusing on its way back to the entrance plane of the
cell. To address this issue, we conduct combined backscattering
and transmission experiments where we image respectively the
scattered intensity pattern at the entrance and exit plane of
the cell onto two identical cameras, aligning the optics so that the
observation windows are the same in both geometries. We
presume that spatial resolution of an event in transmission can
only be obtained when the event occurs near the exit plane of the
cell. Thus, if for a given cell thickness, a spatially distinct event in
transmission can also be resolved in backscattering, we assess
that the critical depth for the visibility of spatially resolved events
is beyond the thickness $L$. Conducting different experiments
varying $L$, we assess that the critical depth corresponds to $L/l^{*} \simeq
7.5$, which is broadly consistent with estimations probing the
resolution depth in speckle imaging techniques~\cite{Zakharov2006}.

Clearly, in both our experiments, the transmission far field and
the backscattering photon correlation imaging experiment, the
visibility of events depends on the thickness of the cell. To obtain
the maximal visibility of the events, while still maintaining $L/l^{*}$
large enough to be within the limit of the multiple scattering
approximations~\cite{KAPLAN1993}, we perform our experiments at age~1 in a cell
with $L = 2$~mm and at age~2 in a cell with $L = 5$~mm. With $l^{*} =
0.44$~mm for age~1 and $l^{*} =
0.73$~mm for age~2 this corresponds to
$L/l^{*} = 4.5$ (age~1) and $L/l^{*} = 6.8$ (age~2). With these $L/l^{*}$ values,
we ensure that all events occurring throughout the depth of the
sample are detected in our backscattering photon correlation
imaging experiments and that the effect of a single event on
$c_{I}\left(t_{w},\tau\right)$ measured in our far field transmission TRC-experiments
is maintained within the same range at both ages; here we recall
that the effect of a single event on $c_{I}\left(t_{w},\tau\right)$ scales with $\left(\xi^{3}/V_{sc}\right)\left(L/
l^{*}\right)^{2}$, where $V_{sc}$ can be modelled by a truncated cone~\cite{Pine1993} and $\xi$ scales
with the bubble size. To be able to directly compare the temporal
evolution of our signals, we perform both experiments simultaneously;
in addition, the diffuse transmittance is measured
continuously from $t_{w} = 0$~s to $t_{w} =  40000$~s, which enables us to
assess the coarsening process for each run. Our CCD detectors
comprise $633 \times 483$ pixels and our linear speckle size is 3 pixels in
the transmission and 1 pixel in the backscattering experiments.
Before each experiment we determine the exact magnification $M$
of the image taken in the backscattering experiment by imaging
a finely marked grid that is placed at the entrance plane of our
cell. Our imaging lens is placed so that $M \cong 1$, such that the
observation area of our experiment is approximately $6.3 \times 4.8
\mathrm{mm}^{2}$. Linear polarizers are placed in front of both cameras so
that only light with a polarization perpendicular to that of the
incident beam is detected; this is of particular importance in
backscattering, where we thereby eliminate the shortest scattered
paths and favour the detection of long, deeply probing paths.
The temporal evolutions of the speckle pattern are recorded at
a frequency of 50 frames/s and 25 frames/s in transmission and
backscattering, respectively, where we adjust the camera exposure
time between 0.1--0.3~ms depending on the average intensity
level. The speckle patterns are subsequently processed to obtain
the temporal evolution of both the spatially averaged signal
$c_{I}\left(t_{w},\tau\right)$ from the transmission experiments and the spatially
resolved signals $c^{\left(s\right)}_{I}\left(t_{w},\tau\right)$ that we present in dynamical activity
maps. While $c_{I}\left(t_{w},\tau\right)$ is processed at different lag-times, we
analyze $c^{\left(s\right)}_{I}\left(t_{w},\tau\right)$ for the shortest accessible experimental lag-time,
$\tau = 0.04$~s, where we use the correction scheme outlined in section~IV\ C of ref.~\cite{Duri2005} to correct for the measurement noise. The size of
the region of interest over which we take the pixel average is
chosen to be $0.1 \times 0.1 \mathrm{mm}^{2}$, which corresponds approximately to
a bubble diameter at both ages.

\section{Results and discussion}

\begin{figure}[tbp]
\begin{center}
\scalebox{1.1}{\includegraphics{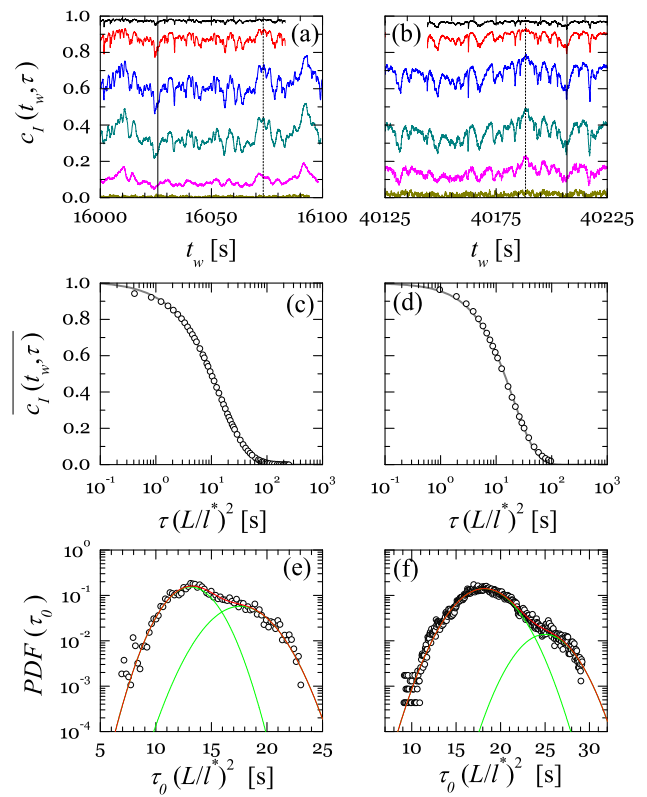}}
\end{center}
\caption{Panel displaying dynamical characteristics obtained from the far
field transmission experiments at age 1 (left) and age 2 (right). (a) and (b):
time resolved degree of correlation as a function of $t_{w}$ calculated for
different lag times (a) from top to bottom: $\tau = 0.04, 0.2, 0.68, 1.6, 4, 12$ s
and (b) from top to bottom: $\tau = 0.04, 0.12, 0.32, 0.8, 1.6, 4$ s. The solid
and dotted vertical lines denote the moments at which the dynamical
activity maps shown in Fig.~\ref{fig:dams} were acquired; the solid lines correspond to
Fig.~\ref{fig:dams}(a) and (c), respectively and the dotted line to Fig.~\ref{fig:dams}(b) and (d). (c)
and (d): time averaged correlation function as a function of lag time; to
account for changes in the experimental geometry $\tau$ is normalised by $\left( L/
l^{*}\right)^{2}$. The continuous lines present the results of exponential fits to the
data. (e) and (f): probability density function of the `instantaneous' decay
time $\tau_{0} \left(t_{w}\right)$ obtained by fitting the age dependent two-time correlation
function to eqn (5). The continuous red lines present the sum of two
Gaussian distributions (green dashed lines) that are used to fit the data.}
\label{fig:trctraces}
\end{figure}

Typical examples of the temporal evolution of the instantaneous
degree of correlation obtained in our transmission far field
experiments are shown for age~1 in Fig.~\ref{fig:trctraces}(a) and for age~2 in
Fig.~\ref{fig:trctraces}(b). The instantaneous degree of correlation systematically
decreases with increasing lag-time, indicating that at any moment
in time the system exhibits dynamical activity. 
Indeed, the
'instantaneous' two-time correlation function can be read as
a vertical cut through the data shown in Fig.~\ref{fig:trctraces}(a) and (b); the
sporadic decreases in $c_{I}\left(t_{w},\tau\right)$ observed at both ages thus effectively
indicate that the dynamical activity of our foams intermittently
increases. Despite this variation, the overall time
averaged correlation function $\overline{c_{I}\left(t_{w},\tau\right)}$ exhibits a simple exponential
decay, as shown in Fig.~\ref{fig:trctraces}(c) and~\ref{fig:trctraces}(d) for age~1 and age~2,
respectively. Such exponential decay is consistent with previous
experiments~\cite{DURIAN1991, Gopal1995,Durian1991PRA} and in principle would be consistent with the
model described above, where the random intermittent restructuring
of volume elements causes a decay of the correlation
function that is indistinguishable to the one of continuously
diffusing particles. However, as we will show, the decay of the
mean correlation function is determined by at least two
dynamical processes, such that this model does not fully apply
for the description of the time averaged correlation function.
Indeed, the inspection of the temporal evolution in the dynamical
activity maps, shown as movie 1 (age~1) and movie 2 (age~2) in the
ESI~\cite{footnote3Dfoams} 
reveals that the events mostly occur one at a time in the
observational volume of our experiment. Moreover, the movies
clearly reveal that the local restructuring of a volume element is
a rather slow process. Indeed, a characterization of the event
duration by calculating the temporal correlation of the dynamical
activity at a given location reveals that the events typically
lasts $\tau_{d} \simeq 1$~s at age 1 and $\tau_{d} \simeq 1.8$~s at age 2.

Between events, the dynamical activity maps reveal that
$c^{\left(s\right)}_{I}\left(t_{w},\tau\right)$ is relatively homogeneous in space, as shown for both
ages in Fig.~\ref{fig:dams}(b) and Fig.~\ref{fig:dams}(d). The direct comparison of the
results obtained in our photon correlation imaging experiment to
those obtained in the simultaneously performed TRC-experiments
shows that the fluctuations in the TRC-traces are effectively
caused by single events of finite duration. However, during
the event-free periods the spatially averaged dynamics probed in
our transmission experiments is not suppressed. This can be seen
for instance at the moments that are marked by a dotted line in
Fig.~\ref{fig:trctraces}(a) and Fig.~\ref{fig:trctraces}(b), which are the moments at which the
dynamical activity maps shown in Fig.~\ref{fig:dams}(a) and (c) were
acquired. Clearly, $c_{I}\left(t_{w},\tau\right)$ still fully decays during these event-free
periods, which indicates that another dynamical process
contributes to the dephasing of the scattered light. As this
process does not lead to significant fluctuations in the degree of
correlation, neither in space nor in time, we assume that this
dynamics is quasi-continuous and occurs everywhere within the
sample. Our findings are consistent with the findings obtained in
recent speckle visibility experiments, where both the finite
duration of the event and the existence of an addition dynamical
process were detected as well~\cite{Gittings2008}. However, in contrast to the
speckle visibility technique, the TRC-technique allows us to
assess the full lag-time dependence of the different dynamical
processes contributing to the dephasing of the scattered light. As
both the duration of the events and the event-free periods
between events are rather long compared to the experimental
decay time of the time-averaged intensity correlation function,
we can attempt to isolate the contribution of the quasi-continuous
process by fitting the two-time correlation function with
\begin{equation}
c_{I}\left(t_{w},\tau\right)= \frac{1}{\beta\left(t_{w}\right)}\exp \left\{-\left(\frac{\tau}{\tau_{0}\left(t_{w}\right)} \right)^{p\left(t_{w}\right)}\right\}
\label{eq:twotimefit}
\end{equation}where we allow the intercept $1/\beta\left(t_{w}\right)$, the decay time $\tau_{0}\left(t_{w}\right)$ and the
stretching exponent $p\left(t_{w}\right)$ to vary with $t_{w}$. At both ages we find
that the probability density functions (PDF) of the instantaneous
decay times $\tau_{0}\left(t_{w}\right)$ exhibit the same basic features: at shorter
decay times, the $PDF(\tau_{0}\left(t_{w}\right))$ exhibits a reasonably well-defined
maximum, at larger decay times the $PDF(\tau_{0}\left(t_{w}\right))$ is characterized by
a shoulder, as shown for age~1 in Fig.~\ref{fig:trctraces}(e) and for age~2 in
Fig.~\ref{fig:trctraces}(f), where we normalize the decay times with $(L/l^{*})^{2}$ to
account for the varying experimental conditions. Our results
clearly show that the decay of the mean-correlation function is
the result of two processes with different characteristic times.
Indeed, we can attribute the time scale of the second maximum to
the one characterizing the quasi-continuous process, while the
time scale of the first maximum still constitutes a convolution of
both the restructuring events and the quasi-continuous process.
In order to determine the characteristic time describing the quasicontinuous
process $\tau_{c}$, we fit the PDFs to a pair of Gaussian
distributions that we find to describe the data reasonably well, as
shown by the red continuous line in Fig.~\ref{fig:trctraces}(e) and~(f). From the
resulting fit parameter $\tau_{0,max2}(L/l^{*})^{2}$, we determine $\tau_{c} = 2 \cdot \tau_{0,max2}(L/l^{*})^{2}$ and report the results in Table~\ref{tab:numbers}; for better
comparison to the original model outlined above (see eqn.~\eqref{eq:g1trans} and~\eqref{eq:g1transVsc}) we choose to consistently state decay times that apply to the
decay of the field correlation function rather than the intensity
correlation, which just entails to apply a factor of two to the
decay time of the intensity correlation function. As compared to
the mean decay-time $\tau_{m} = 2 \cdot \tau_{0}(L/l^{*})^{2}$ determined from the
exponential fit to the data shown in Fig.~\ref{fig:trctraces}(c) and (d), $\tau_{c}$ is larger
by a factor of 1.2 and 1.5 for age 1 and age 2 respectively.
Interestingly, such moderate change in the decay time of the
correlation function was also reported for experiments probing
the foam dynamics before and after a large oscillatory shear
strain was applied~\cite{Gopal1995}. Presuming that a large oscillatory shear
strain erases internal stress imbalances, such that the rearrangement
events are temporally suppressed, it is likely that
the residual dynamics observed after the application of
shear corresponds to the quasi-continuous dynamics observed
here between events. This would suggest that the quasi-continuous
process is not directly linked to the release of imbalanced
stresses.

Finally, we note that the age dependence of the quasi-continuous
process does not correspond to the one of the mean
dynamics, as denoted by comparing the ratios $\tau_{m}$(age 2)/$\tau_{m}$(age 1)
and $\tau_{c}$(age 2)/$\tau_{c}$(age 1) listed in Table~\ref{tab:numbers}. 
\begin{table}[tbp]
\caption{Comparison of coarsening and dynamical characteristics of
Gillette foam at two different ages. $\overline{t_{w}}$ age of the sample at the middle of
the experiment;  $\left\langle a_{t_{w}}\right\rangle$ mean bubble radius; $d\left\langle a_{t_{w}}\right\rangle/dt$ rate of bubble growth; $d\gamma/dt$ strain rate set by the coarsening process; $\tau_{m}$ characteristic decay
time of the time and spatially averaged field auto-correlation function; $\tau_{c}$
decay time characterizing the quasi-continuous process; $R$ rate of events
per unit volume obtained from counting procedure; $\Gamma_{r}$ restructuring rate
obtained from counting procedure; $\tau_{d}$ duration of intermittent events.}
\footnotesize
\begin{tabular}{llllc}
\hline			
  \ & \ &  Age 1 \T \B & Age 2 & Age 1/Age 2 \\
\hline
  $\overline{t_{w}}$ \T & [min]  & 272  & 663  &  \\
  $\left\langle a_{t_{w}}\right\rangle$ & [$\mu$m] & 50 & 76 &  \\
  $d\left\langle a_{t_{w}}\right\rangle/dt$ & [$\mu$m/s] & $1.05 \times 10^{-3}$ \hspace{0.1in} & $0.88 \times 10^{-3}$ & 1.19  \\
  $d\gamma/dt$ & [s$^{-1}$] & $6.87 \times 10^{-5}$ & $3.40 \times 10^{-5}$ & 2.02  \\
  \\
  $\tau_{m}$ & [s] & 29.9 & 35.0  & \\
  $\tau^{-1}_{m}$ & [s$^{-1}$]  & $33.4 \times 10^{-3}$ & $28.6 \times 10^{-3}$ & 1.17 \\
  $\tau_{c}$ & [s] & 35.2 & 58.1 & \\
  $\tau^{-1}_{c}$ & [s$^{-1}$]  & $28.4 \times 10^{-3}$ & $19.3 \times 10^{-3}$ & 1.47 \\
  \\
  $R$ & [s$^{-1}$mm$^{-3}$] & $1.05 \times 10^{-3}$ & $0.88 \times 10^{-3}$ & 1.19 \\
  $\Gamma_{r}$ & [s$^{-1}$] & $4.52 \times 10^{-4}$ & $2.12 \times 10^{-4}$ & 2.13 \\
  $\tau_{d}$ \B & [s] & 1.0 & 1.8 & \\
\hline  
\normalsize
\label{tab:numbers}
\end{tabular}
\end{table}
This indicates that the
event dynamics contributes to the mean dynamics so that
the age-dependence is altered. Moreover, an inspection of the
correlation between $\tau_{0}\left(t_{w}\right)$ and $p\left(t_{w}\right)$ reveals that the quasicontinuous
process is best described by a slightly compressed
exponential, while during the occurrence of events the decay of
the two time correlation function is described by a slightly
stretched exponential. It is indeed remarkable that this combination
between fast and stretched decays and slow and
compressed decays leads to a simple exponential decay in the
mean correlation function. However, this high degree of
convolution of event dynamics and quasi-continuous dynamics
determining the mean dynamics significantly hinders the characterization
of the rate of topological rearrangements and
consequently their dependence on the coarsening process.

By contrast, the characterization of the event dynamics is
rather straightforward using the results of our photon correlation
imaging experiments. Indeed, the temporal evolution of the
events can simply be followed in movies like the one shown in the
ESI~\cite{footnote3Dfoams}. Just counting the events allows us to determine the event
rate per unit volume $R$ reported in Table~\ref{tab:numbers}, where we estimate the
size of our observational volume as the product of the size of the
field of view of our camera and the thickness of the cell.
Clearly, the event rate within a given volume is significantly larger at age 1
than at age 2. However, to evaluate $\Gamma_{r}$, the rate at which a volume
element is effectively restructured, we need to account for the
number of possible rearrangement zones within our observation
volume, which scales with the age-dependent size of the rearranged
zone $\xi$. An analysis of the event size, by calculating the
spatial correlation of $c^{\left(s\right)}_{I}\left(t_{w},\tau\right)$ introduced as $\tilde{G}_{4}\left(\Delta s,\tau\right)$ in ref.~\cite{Duri2009},
reveals that $\xi$ scales with the bubble size and extends over 10
bubble diameters, in agreement with previous assessments~\cite{DURIAN1991,Durian1991PRA,Gittings2008}.
The large spatial extent and long durations of the events
observed in our experiments are consistent with the events probed
at the foam surface, where we observe that a series of
neighbour-switching events, T1-events~\cite{Weaire1999}, rearranges a zone
extending over $\simeq 4$ bubble diameters. The even larger size of the
event observed in diffusing wave spectroscopy can be understood
as that the event typically contains both a central volume
element, in which the bubbles topologically reconfigure, and
a halo of bubbles that respond elastically without rearranging,
such that the zone that effectively contributes to the dephasing of
light extends over $\simeq 10$ bubble diameters~\cite{DURIAN1991,Gopal1995,Gittings2008}.

To evaluate the restructuring rate, we thus assume that the size
of the effectively topological rearranged zone is $\xi = 4 \left\langle a_{t_{w}}\right\rangle$. As
shown in Table~\ref{tab:numbers}, $\Gamma_{r}= R \cdot \frac{4}{3} \pi \left(4\left\langle a_{t_{w}}\right\rangle\right)^{3}$ decreases by
approximately a factor of 2.1 from age 1 to age 2. To assess the
relation between the topological rearrangements and the coarsening
process, we evaluate the strain rate set by the coarsening
process as $d\gamma/dt \left(\Delta V/V\right)/\Delta t$, with $V$ and $\Delta V / \Delta t$ respectively the
bubble volume and the change of the bubble volume per unit
time at a given foam age. As shown in Table~\ref{tab:numbers}, we find that the
ratio of the age dependent strain rates [($d\gamma/dt$(age 1))/($d\gamma/dt$(age
2))= 2.02] correspond, within the experimental error, to the ratio
of the restructuring rates [$\Gamma_{r}$(age 1)/$\Gamma_{r}$(age 2) = 2.1]. This
strongly suggests that the events are a direct result of the
coarsening induced strain between events. Estimating this critical
strain as $\gamma_{y} = (d\gamma/dt)/\Gamma_{r}$, we find that $\gamma_{y} \approx 0.16 \pm 0.01$ in
reasonable agreement with the critical strain measured for our
foams in experiments probing the macroscopic mechanical
properties. This correlation between strain rate and event rate is
not evident when considering the decay rates of the mean
correlation function $\tau^{-1}_{m}$, which, following eqn.~\eqref{eq:g1trans} $\left[\tau^{-1}_{m} = R \cdot \tfrac{4}{3}\pi \xi^{3} \right]$, should effectively correspond to the restructuring rates if the
dynamic light scattering signal would be determined by the
random restructuring of volume elements only. Here the ratio
$\tau^{-1}_{m}$(age 1)/$\tau^{-1}_{m}$(age 2) would appear to correlate to the ratio of the
bubble growth rather than to that of the strain rates, as shown in
Table~\ref{tab:numbers}. Indeed, the convolution of event-frequency, event
duration and quasi-continuous dynamics that all exhibit
a different age-dependence leads to an age-dependence in the
mean-dynamics that is unsuitable for establishing relations
between the coarsening process and topological rearrangements.

To address the question whether not the disappearance of
bubbles, T2-events~\cite{Weaire1999}, rather than the mean strain set by the
coarsening process is at the origin of the mechanical instabilities
that lead to the intermittent rearrangements observed, we estimate
the number of bubbles disappearing within a time interval $\Delta t$ as
\begin{equation}
N_{b}\left(t_{w}\right)-N_{b}\left(t_{w}+\Delta t\right) = \frac{3}{4\pi}\left( \frac{1}{\left\langle a\left(t_{w}\right)\right\rangle^{3}} - \frac{1}{\left\langle a \left(t_{w}+\Delta t\right)\right\rangle^{3}}\right)
\end{equation}
where $N_{b}\left(t_{w}\right)$ and $N_{b}\left(t_{w}+\Delta t\right)$ are respectively the number of
bubbles per unit volume at an age of $t_{w}$ and $t_{w}+\Delta t$. This estimate
yields a rate for T2-events per unit volume that exceeds the rate
of events per unit volume $R$ by a factor of $\cong 10$, such that it
appears unlikely that T2-events are at the actual origin of the
events observed. Finally, we note that single T2-events are small
and involve the motion of only a few plateau-borders, which
have been previously identified to be the principal scattering sites
in dry foams~\cite{Vera2001}. T2-events are thus unlikely to be resolved in our
experiments, as the resolution is limited by the number of paths
rearranged by an event divided by the total number of paths
contributing to $c^{\left(s\right)}_{I}\left(t_{w},\tau\right)$ of a given ROI.

\begin{figure}[tbp]
\begin{center}
\scalebox{1.1}{\includegraphics{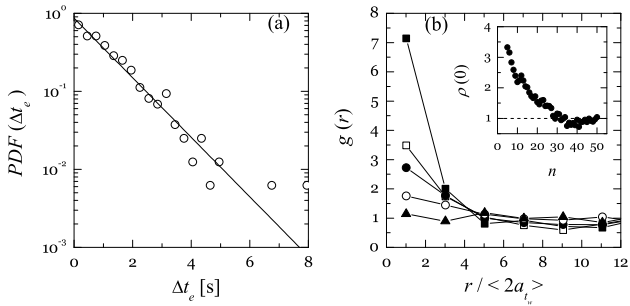}}
\end{center}
\caption{(a) Probability density function of time delay between successive
events. The solid line denotes the exponential behaviour expected for the
random occurrence of independent events. (b) Experimentally determined
probability to find the $(i + n)^{th}$ event observed within our experimental
volume at a distance $r$ from the $i^{th}$ event normalized by the
probability expected for randomly distributed events, $\mathrm{g(r)} = PDF(r)_{experiment} / PDF(r)_{random}$. As $\mathrm{g}(r)$ varies smoothly with increasing $n$, $\mathrm{g}(r)$
obtained for $n =$ 1--5 (filled squares), $n =$ 6--10 (open squares), $n =$ 11--15
(filled circles), $n =$ 16--20 (open circles), and $n =$ 36--40 (triangles) are
averaged to reduce the statistical fluctuations. Inset: Probability that the
$(i + n)^{th}$ event observed within our observational volume is located within
the topologically rearranged zone of the $i^{th}$ event divided by the expectation
value for randomly distributed events.}
\label{fig:gr}
\end{figure}

To further progress in our understanding of the topological
rearrangements, we characterize the temporal and spatial
distribution of the events exploiting the temporally resolved
dynamical activity maps. We assess the time at which an event
occurs as the moment where the dynamical activity is maximal
during an event. At this moment, the location of the event is
determined as the centre of the high dynamical activity zones, as
those shown in Fig.~2(a) and (c). The results obtained from our
analysis of the experiment at age 1 are shown in Fig.~\ref{fig:gr}; qualitatively
similar results are obtained for the experiment at age 2,
but the quality of the data is somewhat lower as the number of
events probed over the duration of our experiment is here lower
than at age 1. 
In agreement with previous results~\cite{Gittings2008}, we find that
the time delay between successive events $\Delta t_{e}$ follows an exponential
distribution, as shown in Fig.~\ref{fig:gr}(a). This is consistent with
the Poissonian statistics expected for the random occurrence of
independent events. To characterize the distribution of the events
in space we determine the distance, $r$, between the $i^{th}$ and $(i + n)^{th}$
event and calculate the probability density function of $r$,
$PDF(r)_{experiment}$, for a given $n$. To account for the limited
geometry of our experiment we determine the PDF expected for
a random distribution of events in space $PDF(r)_{random}$ using
simulations, where we implement the experimental boundary
conditions. By normalising $PDF(r)_{experiment}$  with $PDF(r)_{random}$,
we obtain the radial distribution function $g(r) = PDF(r)_{experiment}/
PDF(r)_{random}$, where we expect $\mathrm{g}(r) = 1$ at all $r$ for a random
distribution of events in space. Instead, we find that $\mathrm{g}(r) > 1$ for
small $r$ and $n < 30$, as shown in Fig.~\ref{fig:gr}(b); as the decay of $\mathrm{g}(r)$
varies only slowly with increasing $n$ we have here averaged each
$\mathrm{g}(r)$ over a few values of $n$ to reduce the statistical fluctuations.
The increased probability at small $r$ indicates that the foam
maintains a spatial ``memory'' of its dynamics; the occurrence of
an event at a given location is thus correlated to previous rearrangements
although no temporal correlation is observed. The
intercept of $\mathrm{g}(r)$ decreases with increasing $n$, indicating that the
structural memory effect weakens with time. Moreover, we note
that $\mathrm{g}(r)$ decreases to the background level at $r$ corresponding to
4--5 bubble diameters; this indicates that the actual centre of an
event is likely to be located within the topologically rearranged
zone of a previous event. To quantify the time over which events
are more likely to occur within such zone, we integrate $\mathrm{g}(r)$ for $r$
up to 4 bubble diameters. The integrated value, $\rho(0)$, compares
the experimentally found event probability within an already
rearranged zone of size $r = 4$ bubble diameters to that of
randomly distributed events. It decreases slowly with $n$, reaching
unity at $n \cong 30$, as shown in the inset of Fig.~\ref{fig:gr}(b), where we have
used a five-point moving average to smooth the data. Corresponding
to $n \cong 30$, the spatial memory time of an event would
thus extend up to $\approx 30$~s; using this time to evaluate the mean
strain set by the coarsening process we find a strain that is 2
orders of magnitudes lower than $\gamma_{y}$.

These findings suggest that a fragile bubble configuration is
likely to still be fragile after reconfiguration. Indeed, if we
consider that the coarsening process leads to imbalances in the
stress-configuration, an event will be set off where the excess
stress exceeds the local yield stress. The reconfiguration process
will proceed until the excess stress is again below the yield stress;
this, however, implies that the excess stress of the recently
rearranged zone is just below the yield stress. As the coarsening
process continuously perturbs the stress configurations
throughout the system, the likelihood that the yield stress within
this zone is soon going to be exceeded again is then high,
consistent with the observed behaviour. An increased probability
to find an event within the already rearranged zone is also
enclosed in the fact that the topological rearrangement leads to
an elastic deformation of the surroundings. Indeed, this elastic
response is likely to further contribute to the unfavourable
stress balance within the just rearranged zone. These conditions
denote that coarsening induced rearrangements are fairly inefficient
in relaxing unbalanced stress configurations and thus
suppressing rearrangements. This contrasts with the effect of
imposing large macroscopic oscillatory strains, which has been
shown to efficiently suppress the occurrence of rearrangement
events~\cite{Gopal1995,Cohen-Addad2001}. Indeed, imposing a strain that significantly exceeds
the yield strain leads to a reconfiguration of the ensemble of the
bubbles; both the initiation and cessation of bubble rearrangements
are here not governed by threshold configurations;
moreover, the topological reconfiguration of the ensemble
eliminates the effect of the elastic deformation, such that
fragility is not preserved.

Finally, we note that a persistence of stress-imbalances within
a given region has been recently inferred from experiments
probing the spatial heterogeneities in the dynamics of
compressed microgel systems~\cite{Sessoms2009}. There it was observed that well-localized
zones of high dynamical activity coexisted with zones of
low dynamical activity and that the high dynamical activity
within a zone persisted in time, which indicates that rearrangements
not necessarily lead to a balanced stress configuration.

To some extent this behaviour is reminiscent of dynamical
facilitation in glasses, where an immobile zone can only become
mobile when the neighbouring zone is already mobile~\cite{Garrahan2002}. As
a direct consequence of the highly cooperative nature in the
dynamics of glassy systems the onset of mobility in a given zone
correlates here to the mobility of the neighbouring zone. In
foams, the onset of mobility in a given zone correlates to previous
mobility within the same zone. One may speculate that both
correlations are related to the actual origin of the dynamics in
glasses and deeply jammed systems. While the thermally-driven
dynamics in glasses imposes cooperativity, i.e. fluidization of
a neighbouring region, rearrangements driven by unbalanced
stresses in deeply jammed systems increase the `degree of
jamming' in the surroundings, thereby inducing an imbalance of
internal stresses at the location that has just been rearranged.

\section{Conclusions}
In conclusion, we have investigated spatial and temporal
heterogeneities in the dynamics of a coarsening foam at two
different ages, conducting simultaneously experiments in the
transmission and backscattering geometry of diffusing wave
spectroscopy. In transmission, we performed the experiments in
the traditional far field limit, while we used the photon correlation
imaging configuration~\cite{Duri2009} in the backscattering experiments.
The combination of these experiments allows us to critically
address different contributions to the dynamic light scattering
signal. In agreement with previous experiments~\cite{Gittings2008}, we find that the
foam dynamics is determined by two processes: intermittent
bubble rearrangements of finite duration and a quasi-continuous
process that is spatially homogeneous. Both processes contribute
to the dynamic light scattering signal, which significantly
complicates the interpretation of the mean signal. For the characterization
of the event dynamics, it becomes evident that the
photon correlation imaging technique is a particularly useful
tool, as it provides the means to assess both the frequency and the
position of the events without the need of modelling the actual
magnitude of the dynamic light scattering signal. These experiments
reveal that the event frequency is directly correlated to the
strain-rate imposed by the coarsening process, which indicates
that the events are triggered by a well defined strain. Our analysis
of the spatial correlation between successive events uncovers that
the events are not randomly distributed in space. Indeed, we find
that the probability to find an event at a location that has been
previously rearranged is significantly higher than expected for
a random spatial distribution of events. This unexpected
behaviour indicates that a locally unstable zone rearranges to
a state that is just below the yield stress, such that the likelihood
that this zone is soon going to be again rearranged is high. Such
persistence might be a general characteristic of stress-driven
dynamics that could be observed in other jammed systems where
the direct contact between the constituents may result in an
unbalanced stress-configuration.

Our findings evoke several interesting questions that could be
addressed in future experiments. In particular, the experiments
probing the effect of a macroscopic applied strain could be used
to explore the fragility of a rearranged zone. One could envision
that a small oscillatory strain would facilitate the secondary
events, such that the events would merge within the typical
duration of a single event; in such a case it is likely that the spatial
distribution of events would appear to be random. Moreover, it
would be interesting to investigate the evolution of the spatial
heterogeneities in the dynamics of a foam that has been subjected to
a large external shear, as this has been done for the mean
dynamics in ref.~\cite{Gopal1995,Cohen-Addad2001}. Such investigations should help to assess
the efficiency of restructuring events in equilibrating stress imbalances and could lead to a better understanding of the quasicontinuous
process whose origin remains unclear.

\section{Acknowledgements} This work has been supported by the Swiss National Science Foundation, the CNES (ACI No.
JC2076) and CNRS (PICS No. 2410). L.C. acknowledges support from the Institut
Universitaire de France. We thank Douglas J. Durian, Reinhard H\"{o}hler and David A. Weitz
for fruitful discussions.

\bibliography{3Dfoambib2}

\begin{thebibliography}{31}
\expandafter\ifx\csname natexlab\endcsname\relax\def\natexlab#1{#1}\fi
\expandafter\ifx\csname bibnamefont\endcsname\relax
  \def\bibnamefont#1{#1}\fi
\expandafter\ifx\csname bibfnamefont\endcsname\relax
  \def\bibfnamefont#1{#1}\fi
\expandafter\ifx\csname citenamefont\endcsname\relax
  \def\citenamefont#1{#1}\fi
\expandafter\ifx\csname url\endcsname\relax
  \def\url#1{\texttt{#1}}\fi
\expandafter\ifx\csname urlprefix\endcsname\relax\def\urlprefix{URL }\fi
\providecommand{\bibinfo}[2]{#2}
\providecommand{\eprint}[2][]{\url{#2}}

\bibitem[{\citenamefont{Weaire and Hutzler}(1999)}]{Weaire1999}
\bibinfo{author}{\bibfnamefont{D.}~\bibnamefont{Weaire}} \bibnamefont{and}
  \bibinfo{author}{\bibfnamefont{S.}~\bibnamefont{Hutzler}},
  \emph{\bibinfo{title}{The Physics of Foams}} (\bibinfo{publisher}{Claredon
  Press}, \bibinfo{year}{1999}).

\bibitem[{\citenamefont{Durian et~al.}(1991{\natexlab{a}})\citenamefont{Durian,
  Weitz, and Pine}}]{DURIAN1991}
\bibinfo{author}{\bibfnamefont{D.~J.} \bibnamefont{Durian}},
  \bibinfo{author}{\bibfnamefont{D.~A.} \bibnamefont{Weitz}}, \bibnamefont{and}
  \bibinfo{author}{\bibfnamefont{D.~J.} \bibnamefont{Pine}},
  \bibinfo{journal}{Science} \textbf{\bibinfo{volume}{252}},
  \bibinfo{pages}{686} (\bibinfo{year}{1991}{\natexlab{a}}).

\bibitem[{\citenamefont{Sollich et~al.}(1997)\citenamefont{Sollich, Lequeux,
  Hébraud, and Cates}}]{Sollich1997}
\bibinfo{author}{\bibfnamefont{P.}~\bibnamefont{Sollich}},
  \bibinfo{author}{\bibfnamefont{F.}~\bibnamefont{Lequeux}},
  \bibinfo{author}{\bibfnamefont{P.}~\bibnamefont{Hébraud}}, \bibnamefont{and}
  \bibinfo{author}{\bibfnamefont{M.~E.} \bibnamefont{Cates}},
  \bibinfo{journal}{Phys. Rev. Lett.} \textbf{\bibinfo{volume}{78}},
  \bibinfo{pages}{2020} (\bibinfo{year}{1997}).

\bibitem[{\citenamefont{Cipelletti et~al.}(2000)\citenamefont{Cipelletti,
  Manley, Ball, and Weitz}}]{Cipelletti2000}
\bibinfo{author}{\bibfnamefont{L.}~\bibnamefont{Cipelletti}},
  \bibinfo{author}{\bibfnamefont{S.}~\bibnamefont{Manley}},
  \bibinfo{author}{\bibfnamefont{R.~C.} \bibnamefont{Ball}}, \bibnamefont{and}
  \bibinfo{author}{\bibfnamefont{D.~A.} \bibnamefont{Weitz}},
  \bibinfo{journal}{Phys. Rev. Lett.} \textbf{\bibinfo{volume}{84}},
  \bibinfo{pages}{2275} (\bibinfo{year}{2000}).

\bibitem[{\citenamefont{Sessoms et~al.}(2009)\citenamefont{Sessoms,
  Bischofberger, Cipelletti, and Trappe}}]{Sessoms2009}
\bibinfo{author}{\bibfnamefont{D.~A.} \bibnamefont{Sessoms}},
  \bibinfo{author}{\bibfnamefont{I.}~\bibnamefont{Bischofberger}},
  \bibinfo{author}{\bibfnamefont{L.}~\bibnamefont{Cipelletti}},
  \bibnamefont{and} \bibinfo{author}{\bibfnamefont{V.}~\bibnamefont{Trappe}},
  \bibinfo{journal}{Philosophical Transactions of the Royal Society A}
  \textbf{\bibinfo{volume}{367}}, \bibinfo{pages}{5013} (\bibinfo{year}{2009}).

\bibitem[{\citenamefont{Ramos and Cipelletti}(2005)}]{Ramos2005}
\bibinfo{author}{\bibfnamefont{L.}~\bibnamefont{Ramos}} \bibnamefont{and}
  \bibinfo{author}{\bibfnamefont{L.}~\bibnamefont{Cipelletti}},
  \bibinfo{journal}{Phys. Rev. Lett.} \textbf{\bibinfo{volume}{94}},
  \bibinfo{pages}{158301} (\bibinfo{year}{2005}).

\bibitem[{\citenamefont{Cohen-Addad and H\"{o}hler}(2001)}]{Cohen-Addad2001}
\bibinfo{author}{\bibfnamefont{S.}~\bibnamefont{Cohen-Addad}} \bibnamefont{and}
  \bibinfo{author}{\bibfnamefont{R.}~\bibnamefont{H\"{o}hler}},
  \bibinfo{journal}{Phys. Rev. Lett.} \textbf{\bibinfo{volume}{86}},
  \bibinfo{pages}{4700} (\bibinfo{year}{2001}).

\bibitem[{\citenamefont{Gopal and Durian}(1995)}]{Gopal1995}
\bibinfo{author}{\bibfnamefont{A.~D.} \bibnamefont{Gopal}} \bibnamefont{and}
  \bibinfo{author}{\bibfnamefont{D.~J.} \bibnamefont{Durian}},
  \bibinfo{journal}{Phys. Rev. Lett.} \textbf{\bibinfo{volume}{75}},
  \bibinfo{pages}{2610} (\bibinfo{year}{1995}).

\bibitem[{\citenamefont{H\"{o}hler et~al.}(1997)\citenamefont{H\"{o}hler,
  Cohen-Addad, and Hoballah}}]{Hoehler1997}
\bibinfo{author}{\bibfnamefont{R.}~\bibnamefont{H\"{o}hler}},
  \bibinfo{author}{\bibfnamefont{S.}~\bibnamefont{Cohen-Addad}},
  \bibnamefont{and} \bibinfo{author}{\bibfnamefont{H.}~\bibnamefont{Hoballah}},
  \bibinfo{journal}{Phys. Rev. Lett.} \textbf{\bibinfo{volume}{79}},
  \bibinfo{pages}{1154} (\bibinfo{year}{1997}).

\bibitem[{\citenamefont{Earnshaw and Jaafar}(1994)}]{Earnshaw1994}
\bibinfo{author}{\bibfnamefont{J.~C.} \bibnamefont{Earnshaw}} \bibnamefont{and}
  \bibinfo{author}{\bibfnamefont{A.~H.} \bibnamefont{Jaafar}},
  \bibinfo{journal}{Phys. Rev. E} \textbf{\bibinfo{volume}{49}},
  \bibinfo{pages}{5408} (\bibinfo{year}{1994}).

\bibitem[{\citenamefont{Gopal and Durian}(1999)}]{Gopal1999}
\bibinfo{author}{\bibfnamefont{A.~D.} \bibnamefont{Gopal}} \bibnamefont{and}
  \bibinfo{author}{\bibfnamefont{D.~J.} \bibnamefont{Durian}},
  \bibinfo{journal}{Journal of Colloid and Interface Science}
  \textbf{\bibinfo{volume}{213}}, \bibinfo{pages}{169} (\bibinfo{year}{1999}),
  ISSN \bibinfo{issn}{0021-9797}.

\bibitem[{\citenamefont{Gopal and Durian}(1997)}]{Gopal1997}
\bibinfo{author}{\bibfnamefont{A.~D.} \bibnamefont{Gopal}} \bibnamefont{and}
  \bibinfo{author}{\bibfnamefont{D.~J.} \bibnamefont{Durian}},
  \bibinfo{journal}{J. Opt. Soc. Am. A} \textbf{\bibinfo{volume}{14}},
  \bibinfo{pages}{150} (\bibinfo{year}{1997}).

\bibitem[{\citenamefont{Pine and Weitz}(1993)}]{Pine1993}
\bibinfo{author}{\bibfnamefont{D.}~\bibnamefont{Pine}} \bibnamefont{and}
  \bibinfo{author}{\bibfnamefont{D.}~\bibnamefont{Weitz}},
  \emph{\bibinfo{title}{Dynamic Light Scattering}} (\bibinfo{publisher}{Oxford
  University Press}, \bibinfo{year}{1993}), chap.
  \bibinfo{chapter}{Diffusing-wave spectroscopy}.

\bibitem[{\citenamefont{Maret}(1997)}]{Maret1997}
\bibinfo{author}{\bibfnamefont{G.}~\bibnamefont{Maret}},
  \bibinfo{journal}{Current Opinion in Colloid \& Interface Science}
  \textbf{\bibinfo{volume}{2}}, \bibinfo{pages}{251} (\bibinfo{year}{1997}).

\bibitem[{\citenamefont{Gittings and Durian}(2008)}]{Gittings2008}
\bibinfo{author}{\bibfnamefont{A.~S.} \bibnamefont{Gittings}} \bibnamefont{and}
  \bibinfo{author}{\bibfnamefont{D.~J.} \bibnamefont{Durian}},
  \bibinfo{journal}{Phys. Rev. E} \textbf{\bibinfo{volume}{78}},
  \bibinfo{pages}{066313} (\bibinfo{year}{2008}).

\bibitem[{\citenamefont{Duri et~al.}(2009)\citenamefont{Duri, Sessoms, Trappe,
  and Cipelletti}}]{Duri2009}
\bibinfo{author}{\bibfnamefont{A.}~\bibnamefont{Duri}},
  \bibinfo{author}{\bibfnamefont{D.~A.} \bibnamefont{Sessoms}},
  \bibinfo{author}{\bibfnamefont{V.}~\bibnamefont{Trappe}}, \bibnamefont{and}
  \bibinfo{author}{\bibfnamefont{L.}~\bibnamefont{Cipelletti}},
  \bibinfo{journal}{Phys. Rev. Lett.} \textbf{\bibinfo{volume}{102}},
  \bibinfo{pages}{085702} (\bibinfo{year}{2009}).

\bibitem[{\citenamefont{Cipelletti et~al.}(2003)\citenamefont{Cipelletti,
  Bissig, Trappe, Ballesta, and Mazoyer}}]{Cipelletti2003}
\bibinfo{author}{\bibfnamefont{L.}~\bibnamefont{Cipelletti}},
  \bibinfo{author}{\bibfnamefont{H.}~\bibnamefont{Bissig}},
  \bibinfo{author}{\bibfnamefont{V.}~\bibnamefont{Trappe}},
  \bibinfo{author}{\bibfnamefont{P.}~\bibnamefont{Ballesta}}, \bibnamefont{and}
  \bibinfo{author}{\bibfnamefont{S.}~\bibnamefont{Mazoyer}},
  \bibinfo{journal}{Journal of Physics: Condensed Matter}
  \textbf{\bibinfo{volume}{15}}, \bibinfo{pages}{S257} (\bibinfo{year}{2003}).

\bibitem[{\citenamefont{Duri et~al.}(2005)\citenamefont{Duri, Bissig, Trappe,
  and Cipelletti}}]{Duri2005}
\bibinfo{author}{\bibfnamefont{A.}~\bibnamefont{Duri}},
  \bibinfo{author}{\bibfnamefont{H.}~\bibnamefont{Bissig}},
  \bibinfo{author}{\bibfnamefont{V.}~\bibnamefont{Trappe}}, \bibnamefont{and}
  \bibinfo{author}{\bibfnamefont{L.}~\bibnamefont{Cipelletti}},
  \bibinfo{journal}{Phys. Rev. E} \textbf{\bibinfo{volume}{72}},
  \bibinfo{pages}{051401} (\bibinfo{year}{2005}).

\bibitem[{\citenamefont{Durian et~al.}(1991{\natexlab{b}})\citenamefont{Durian,
  Weitz, and Pine}}]{Durian1991PRA}
\bibinfo{author}{\bibfnamefont{D.~J.} \bibnamefont{Durian}},
  \bibinfo{author}{\bibfnamefont{D.~A.} \bibnamefont{Weitz}}, \bibnamefont{and}
  \bibinfo{author}{\bibfnamefont{D.~J.} \bibnamefont{Pine}},
  \bibinfo{journal}{Phys. Rev. A} \textbf{\bibinfo{volume}{44}},
  \bibinfo{pages}{R7902} (\bibinfo{year}{1991}{\natexlab{b}}).

\bibitem[{\citenamefont{Cohen-Addad et~al.}(1998)\citenamefont{Cohen-Addad,
  Hoballah, and H\"{o}hler}}]{Cohen-Addad1998}
\bibinfo{author}{\bibfnamefont{S.}~\bibnamefont{Cohen-Addad}},
  \bibinfo{author}{\bibfnamefont{H.}~\bibnamefont{Hoballah}}, \bibnamefont{and}
  \bibinfo{author}{\bibfnamefont{R.}~\bibnamefont{H\"{o}hler}},
  \bibinfo{journal}{Phys. Rev. E} \textbf{\bibinfo{volume}{57}},
  \bibinfo{pages}{6897} (\bibinfo{year}{1998}).

\bibitem[{\citenamefont{H\"{o}hler et~al.}(1999)\citenamefont{H\"{o}hler,
  Cohen-Addad, and Asnacios}}]{HohlerEPL1999}
\bibinfo{author}{\bibfnamefont{R.}~\bibnamefont{H\"{o}hler}},
  \bibinfo{author}{\bibfnamefont{S.}~\bibnamefont{Cohen-Addad}},
  \bibnamefont{and} \bibinfo{author}{\bibfnamefont{A.}~\bibnamefont{Asnacios}},
  \bibinfo{journal}{EPL (Europhysics Letters)} \textbf{\bibinfo{volume}{48}},
  \bibinfo{pages}{93} (\bibinfo{year}{1999}).

\bibitem[{\citenamefont{Cohen-Addad et~al.}(2004)\citenamefont{Cohen-Addad,
  H\"{o}hler, and Khidas}}]{Cohen-Addad2004}
\bibinfo{author}{\bibfnamefont{S.}~\bibnamefont{Cohen-Addad}},
  \bibinfo{author}{\bibfnamefont{R.}~\bibnamefont{H\"{o}hler}},
  \bibnamefont{and} \bibinfo{author}{\bibfnamefont{Y.}~\bibnamefont{Khidas}},
  \bibinfo{journal}{Phys. Rev. Lett.} \textbf{\bibinfo{volume}{93}},
  \bibinfo{pages}{028302} (\bibinfo{year}{2004}).

\bibitem[{\citenamefont{Briers}(2007)}]{Briers2007}
\bibinfo{author}{\bibfnamefont{J.~D.} \bibnamefont{Briers}},
  \bibinfo{journal}{Optica Applicata} \textbf{\bibinfo{volume}{37}},
  \bibinfo{pages}{139} (\bibinfo{year}{2007}).

\bibitem[{\citenamefont{Zakharov and Scheffold}(2010)}]{Zakharov2010}
\bibinfo{author}{\bibfnamefont{P.}~\bibnamefont{Zakharov}} \bibnamefont{and}
  \bibinfo{author}{\bibfnamefont{F.}~\bibnamefont{Scheffold}},
  \bibinfo{journal}{Soft Materials} \textbf{\bibinfo{volume}{8}},
  \bibinfo{pages}{102} (\bibinfo{year}{2010}), ISSN \bibinfo{issn}{1539-445X}.

\bibitem[{\citenamefont{Erpelding et~al.}(2008)\citenamefont{Erpelding, Amon,
  and Crassous}}]{Erpelding2008}
\bibinfo{author}{\bibfnamefont{M.}~\bibnamefont{Erpelding}},
  \bibinfo{author}{\bibfnamefont{A.}~\bibnamefont{Amon}}, \bibnamefont{and}
  \bibinfo{author}{\bibfnamefont{J.}~\bibnamefont{Crassous}},
  \bibinfo{journal}{Phys. Rev. E} \textbf{\bibinfo{volume}{78}},
  \bibinfo{pages}{046104} (\bibinfo{year}{2008}).

\bibitem[{\citenamefont{Cerbino and Vailati}(2009)}]{Cerbino2009}
\bibinfo{author}{\bibfnamefont{R.}~\bibnamefont{Cerbino}} \bibnamefont{and}
  \bibinfo{author}{\bibfnamefont{A.}~\bibnamefont{Vailati}},
  \bibinfo{journal}{Current Opinion in Colloid \& Interface Science}
  \textbf{\bibinfo{volume}{14}}, \bibinfo{pages}{416} (\bibinfo{year}{2009}).

\bibitem[{\citenamefont{Zakharov et~al.}(2006)\citenamefont{Zakharov, Volker,
  Buck, Weber, and Scheffold}}]{Zakharov2006}
\bibinfo{author}{\bibfnamefont{P.}~\bibnamefont{Zakharov}},
  \bibinfo{author}{\bibfnamefont{A.}~\bibnamefont{Volker}},
  \bibinfo{author}{\bibfnamefont{A.}~\bibnamefont{Buck}},
  \bibinfo{author}{\bibfnamefont{B.}~\bibnamefont{Weber}}, \bibnamefont{and}
  \bibinfo{author}{\bibfnamefont{F.}~\bibnamefont{Scheffold}},
  \bibinfo{journal}{Optics Letters} \textbf{\bibinfo{volume}{31}},
  \bibinfo{pages}{3465} (\bibinfo{year}{2006}).

\bibitem[{\citenamefont{Kaplan et~al.}(1993)\citenamefont{Kaplan, Kao, Yodh,
  and Pine}}]{KAPLAN1993}
\bibinfo{author}{\bibfnamefont{P.~D.} \bibnamefont{Kaplan}},
  \bibinfo{author}{\bibfnamefont{M.~H.} \bibnamefont{Kao}},
  \bibinfo{author}{\bibfnamefont{A.~G.} \bibnamefont{Yodh}}, \bibnamefont{and}
  \bibinfo{author}{\bibfnamefont{D.~J.} \bibnamefont{Pine}},
  \bibinfo{journal}{Applied Optics} \textbf{\bibinfo{volume}{32}},
  \bibinfo{pages}{3828} (\bibinfo{year}{1993}).

\bibitem[{foo(2010)}]{footnote3Dfoams}
\bibinfo{journal}{Electronic supplementary information (ESI) available: Movies
  showing the evolution of the spatially resolved dynamical activity of
  Gillette shaving foam at age 1 (movie 1) and age 2 (movie 2). The scale bars
  correspond to 1 mm. The observational volumes are respectively $V_{sc} = 67
  \mathrm{mm}^{3}$ and $V_{sc} = 144 \mathrm{mm}^{3}$ for the experiments at
  age 1 and age 2. The number of the dynamical independent zones within the
  observational volume, $V_{sc}/(\tfrac{4}{3} \pi \xi^{3})$, is by a factor of
  1.6 higher in the experiment at age 1 than in the experiment at age 2. See
  DOI: 10.1039/b926873a}  (\bibinfo{year}{2010}),
  \urlprefix\url{http://www.rsc.org/Publishing/Journals/SM/article.asp?doi=b92%
6873a}.

\bibitem[{\citenamefont{Vera et~al.}(2001)\citenamefont{Vera, Saint-Jalmes, and
  Durian}}]{Vera2001}
\bibinfo{author}{\bibfnamefont{M.~U.} \bibnamefont{Vera}},
  \bibinfo{author}{\bibfnamefont{A.}~\bibnamefont{Saint-Jalmes}},
  \bibnamefont{and} \bibinfo{author}{\bibfnamefont{D.~J.}
  \bibnamefont{Durian}}, \bibinfo{journal}{Appl. Opt.}
  \textbf{\bibinfo{volume}{40}}, \bibinfo{pages}{4210} (\bibinfo{year}{2001}).

\bibitem[{\citenamefont{Garrahan and Chandler}(2002)}]{Garrahan2002}
\bibinfo{author}{\bibfnamefont{J.~P.} \bibnamefont{Garrahan}} \bibnamefont{and}
  \bibinfo{author}{\bibfnamefont{D.}~\bibnamefont{Chandler}},
  \bibinfo{journal}{Phys. Rev. Lett.} \textbf{\bibinfo{volume}{89}},
  \bibinfo{pages}{035704} (\bibinfo{year}{2002}).

\end{thebibliography}

\end{document}